# A new approach to services differentiation between mobile terminals of a wireless LAN

*Maher BEN JEMAA*
ReDCAD Research Unit, National School of Engineering of Sfax
BP 1173-3038 Sfax, Tunisia
Maher.benjemaa@enis.rnu.tn

*Maryam KALLEL ZOUARI*
ReDCAD Research Unit, National School of Engineering of Sfax
BP 1173-3038 Sfax, Tunisia
Maryam.kallel@yahoo.fr

Bachar ZOUARI
ReDCAD Research Unit, National School of Engineering of Sfax
BP 1173-3038 Sfax, Tunisia
Bachar.zouari@zouari.net

**Abstract**— This study aims to identify the advantages and disadvantages of several mechanisms for service differentiation in mobile terminals of a wireless LAN to establish a better and more efficient network. According to the analysis of available approaches for the quality of service of the IEEE 802.11 standard, the objective of this paper is to suggest a new method named DF-DCF "Differentiated Frame DCF". DF-DCF can be regarded as an implementation of the algorithm EDF (Earliest Deadline First). The system using DF-DCF is considered as a system with time-dependent priorities. The performance of the suggested method in a Network Simulator (NS) environment allowed its validation through a set of testing and simulation scenarios. Simulation results have shown that the DF-DCF method is better suited for mobile nodes in a wireless communication network.

*Keywords*- Service Differentiation, Wireless LAN, mobility, DCF, DF-DCF, NS.

## I. INTRODUCTION

Recently, Wireless LANs based on IEEE 802.11 standard offers speeds of up to 11 Mbits/s (about 6.5 Mbps in practice). In addition to the extension of wired LANs, WLANs have generated new markets, including public access networks or public hot spots. From the standpoint of standardization, there are currently two standards for WLANs: High Performance Radio LAN (HiperLan) [1, 2] and IEEE 802.11 [3]. However due to the emergence of WiFi products [4, 5] and WiFi5 [6], the IEEE 802.11 standard has recently a major success that continues to grow. We are interested in this work to study the quality of service in this network family. Indeed, access to multimedia content can be achieved only if these networks offer guarantees in terms of delay, jitter or loss rate.

The IEEE 802.11 MAC layer includes a large number of management features such as the frames addressing, frame formatting, error checking, fragmentation and frame reassembly, the management of the terminals mobility (association, reassociation, disassociation) and security services (authentication, desauthentification, privacy). Aside from these management features, one of the features of the 802.11 MAC layer is that it defines two different methods of access to the medium. The first is the Distributed Coordination Function (DCF), which corresponds to an access method similar to that of Ethernet. The DCF was designed to support the transport of asynchronous data and to allow all users wishing to transmit data to have the same opportunity to access support. In the second, the "Point Coordination Function" (PCF), the various data transmissions between the network terminals are managed by a central point of coordination. This is usually located in the access point. The PCF has been designed to enable the transmission of sensitive data. We found that the different proposals for the quality of service introduction in WLANs have some inefficiency. On the one hand, centralized access methods are complex to implement, and on the other hand, completely distributed methods provide an effective differentiated service in the absence of TCP traffic. To address the limitations, we propose in this paper a new mechanism for service differentiation DF-DCF. This mechanism is based on the idea that for any given real-time flow is associated a period beyond which it becomes unmanageable. Based on this fact, we propose to choose the parameters of differentiated services to be applied to the frame also according to the time held in the queue. If this deadline is too large, the frame will simply be abandoned. Since DIFS differentiation is one that provides the best results in terms of differentiation of services, we propose to improve and integrate the period of queuing in the calculation of an instantaneous value for DIFS per frame.

The paper is organized as follows. In section 2, we describe the main extensions to the IEEE 802.11for supporting quality of service. Section 3 is devoted to the description of our proposed approach: DF-DCF to better meet the requirements of different quality of service. A performance analysis and a comparative study with DIFS approach will be presented in Section 4. Finally, a conclusion and perspectives close this paper.

## II. PROPOSED APPROACH: DF-DCF

With the prospect of improvement in the differentiation of services for both the TCP and UDP traffics, we propose a new mechanism for differentiation of services per frame in WLANs named DF-DCF. The method of access to medium in DF-DCF is based on the use of the time for queuing frames in the calculation of instantaneous DIFS values to be applied to these frames of data. The basic idea is to take into account the delay that may have the frames belonging to a flow to calculate their priority [14]. We associate with each frame a time of expiry which is the time when the frame must be transmitted. If the frame is not transmitted before the expiration of its lifetime, it will be eliminated. The service level of the frame will be calculated on the basis of its residual life. This is done through a feature called FSL for 'Frame Service Level". The service differentiation is based primarily on the basis of calculating the level of service by frame: the frame with the smallest deadline is transmitted firstly. In order to implement this level of service in the MAC layer, it is done by calculating a value of DIFS for each frame depending on its level of service FSL. The two following sections describe the two essential components of the proposal: (i) calculate the level of service per frame and (ii) calculate the value of DIFS which corresponds to the class of service to which the frame belongs.

### A. Level of service per frame

$Temax_j$ is the maximum delay allowed for a frame belonging to the class of service j. The level of service per frame, $FSL_j(t)$, calculated at time t is then defined by equation (1).

$$FSL_j(t) = (Temax_j + \tau - t) / Temax_j \quad (1)$$

Where $\tau$ is the moment of arrival of the frame.

Note that the function $FSL_j(t)$ thus defined can be regarded as an implementation of the algorithm EDF (Earliest Deadline First). The system using $FSL_j(t)$ is described as a system with time-dependent priorities [15].

Fig. 1 illustrates how service levels are calculated for each frame. In this example, we consider the respective arrivals at times $t_1$ and $t_2$ of two frames F11 and F21 belonging to two different classes of service 1 and 2 having the constraints of time equal to the maximum $Temax_1$ and $Temax_2 < Temax_1$.

The first frame to process in a system with time dependent services has the highest level of service snapshot (the one whose value $FSL_j(t)$ is minimal). In the example shown in Fig. 1, the frame F11 will be processed primarily compared to F21 if frame processing time is in the interval $[t_1, t_{th}]$ where $t_{th}$ corresponds to the moment when the $FSL_j(t)$ has the same value for both frames ($FSL_1(t_{th}) = FSL_2(t_{th})$) and therefore at the moment when they have the same level of service. It is important to note that during this time interval, frame F11 is chosen despite the fact that it belongs to the lowest priority class. For any time greater than or equal to the moment $t_{th}$, it will be the frame F21 which will be primarily chosen. Note that in such a system, a frame is removed as soon as its function $FSL_j(t)$ reaches zero.

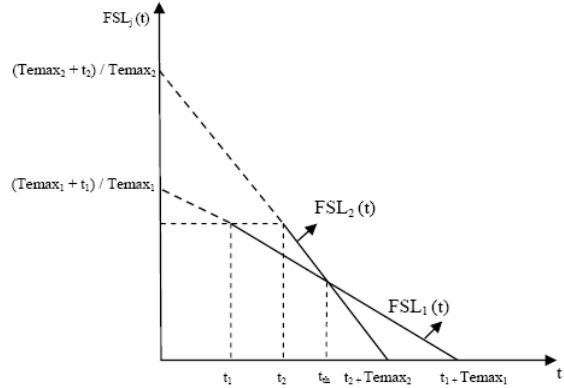

Figure 1. Function level of service per frame - $FSL_j(t)$

### B. Calculating the DIFS

$FSL_j(t)$ function allows to calculate the service level of a frame following the waiting period that has been in the queue. The service classes are differentiated by the lifetime limit for a frame, $Temax_j$. The question we will answer is: how to use this service to obtain a differentiation of services at the MAC layer? To answer this question, we were inspired by the results obtained in [8, 9]. Specifically, we propose to extend the DIFS differentiation by introducing the service level of each frame, $FSL_j(t)$ in calculating the value DIFS, which then becomes instantaneous and per frame. A class of service is defined by the following triplet ($Temax_j$, $DIFS_j^{min}$, $DIFS_j^{max}$) where $DIFS_j^{min}$ and $DIFS_j^{max}$ are the minimum and maximum values DIFS that can take a frame in the same class of service.

$$DIFS_j = SIFS + nbSlotDIFS_j * SlotTime \quad (2)$$

Where $nbSlotDIFS_j$ is the parameter for the differentiation of services. In the standard DCF approach, this parameter is 2.

Our proposal is to calculate the DIFS value for a frame belonging to the $i^{th}$ class of service at time t, where the frame is selected for transmission by the MAC layer as follows:

$DIFS_j(t) = SIFS + (nbSlotDIFS_j^{min} + \lfloor (nbSlotDIFS_j^{max} - nbSlotDIFS_j^{min}) * FSL_j(t) \rfloor) * SlotTime$ (3)

The two parameters $nbSlotDIFS_j^{max}$ and $nbSlotDIFS_j^{min}$ are derived from calculation of $DIFS_j^{max}$ and $DIFS_j^{min}$ such as:

$DIFS_j^{min} = SIFS + nbSlotDIFS_j^{min} * SlotTime$ (4)
$DIFS_j^{max} = SIFS + nbSlotDIFS_j^{max} * SlotTime$ (5)

Equation (5) shows that the value $DIFS_j(t)$ calculated at time t for a frame of class j is assigned by calculating the level of this frame $FSL_j(t)$. Thus, smaller will be the value $FSL_j(t)$ smaller will be the value $DIFS_j(t)$. Hence, more a frame is waiting for transmission, more deadline approaches, and the level will increase accordingly and thus it will have the chance to access the medium when it is supported by the MAC sub-layer. In DF-DCF, the class which is more priority is the one that having the smallest value $Temax_j$ because it has the hard deadline. In case when the two classes have the same value $Temax_j$, the class having the highest priority will be the one with the smallest $DIFS_j^{min}$, then, if still equal, the one with the smallest $DIFS_j^{max}$.

## III. PERFORMANCE EVALUATION OF DF-DCF AND COMPARISON WITH THE EXISTING DIFS APPROACH

To study the performance of DF-DCF, we have extended the implementation of the IEEE 802.11 standard available on NS [16]. We have included the optional use of access methods DF-DCF and DIFS as an alternative to DCF. In this implementation, the capacity of the radio link is 1Mbps. The simulation results obtained using DF-DCF are compared with those obtained by DIFS only. This comparison is made for several possible scenarios. All these scenarios use the same topology: three stations (STA1, STA2 and STA3) are evenly distributed around an access point (AP). These three stations send three streams to a fixed terminal wired connected to the AP, as illustrated by the Figure 2.

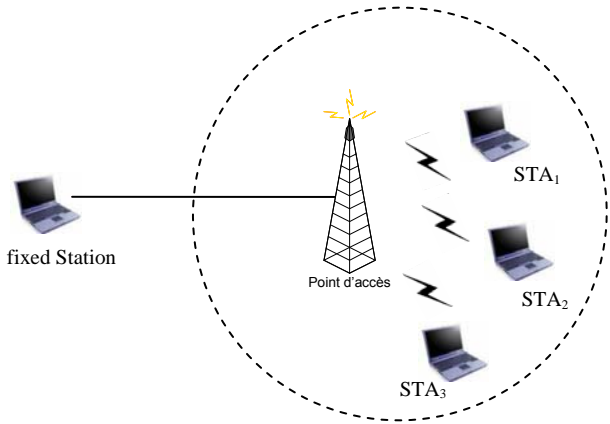

Figure 2. Simulation Topology

The three flows from the three stations will be competing for access to the medium. Each of these three streams will be assigned to a particular class of service. For each scenario, the transmission of three streams begins respectively 50s, 100s and 150s in a simulation whose duration is 250s. To demonstrate the contribution of our approach, its obtained results will be compared to a DIFS differentiation. A class of service will be defined by the triplet ($DIFS_j^{min}$, $DIFS_j^{max}$, $Temax_j$) for DF-DCF and the parameter $DIFS_j$ for DIFS differentiation.

TABLE I. PARAMETERS OF THE SERVICE CLASSES FOR DF-DCF AND DIFS DIFFERENTIATION

|      | DF-DCF | | DIFS |
|------|---------|---------|------|
|      | $DIFS_j^{min}$ / $DIFS_j^{max}$ | $Temax_j$ | $DIFS_j$ |
| **CBR1** | 50µs / 130µs | 150ms | 50µs |
| **CBR2** | 130µs / 210µs | 250ms | 130µs |
| **CBR3** | 210µs / 290µs | 350ms | 210µs |

### A. UDP Traffic

In the first experiment, the three mobiles sent three UDP flows at constant flow: CBR1, CBR2 and CBR3. In this experiment, each of the three streams will saturate the radio link by transmitting at a rate exceeding the capacity of the channel. To do this, each flow sends packets of 2312 bytes every 0.02 s. The highest priority is given to CBR1, a medium priority to CBR2 and the smallest priority to CBR3. The parameters representing the respective priorities of these three streams for each method of access being simulated are illustrated in TABLE I.

(a) Delay using three different flows CBR/UDP: case of DIFS

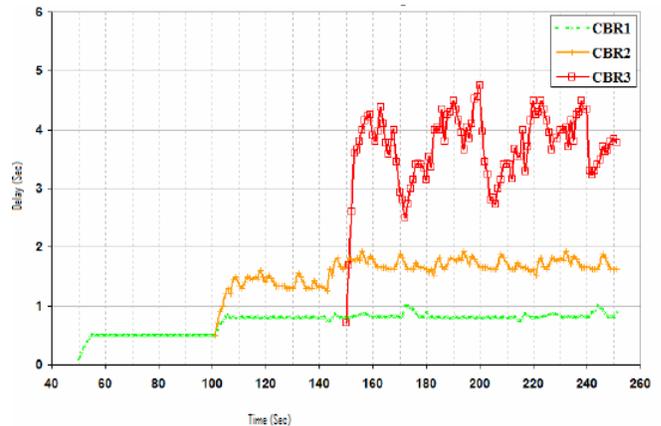

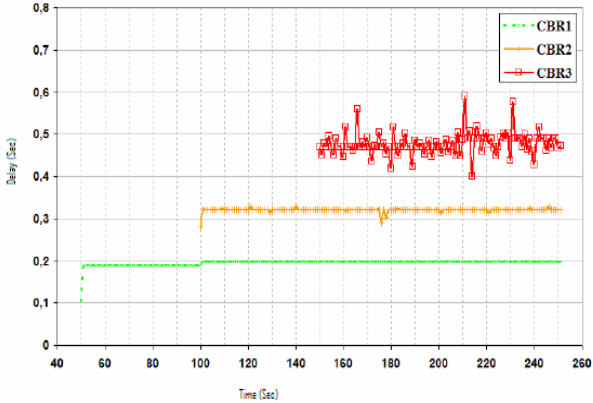

(b) Delay using three different flows CBR/UDP: case of DF-DCF

Figure 3. Delay introduced by the MAC layer: (a) DIFS, (b) DF-DCF.

Figure 3.a/b shows the delays by the three flows. The curves of Figure 3.a/b clearly show that the differentiation of services obtained by using DF-DCF is better than that obtained using DIFS (Figure 3.a).

In addition, the comparison between the two mechanisms for each flow shows that delays obtained with our scheme are lower than those obtained by DIFS. This result was indeed expected because in DF-DCF we try to keep the waiting time in queue of the terminal, less than $Temax_j$. Our system reduces delays by eliminating frames whose lifetime has expired. This allows more frames to be transmitted more quickly.

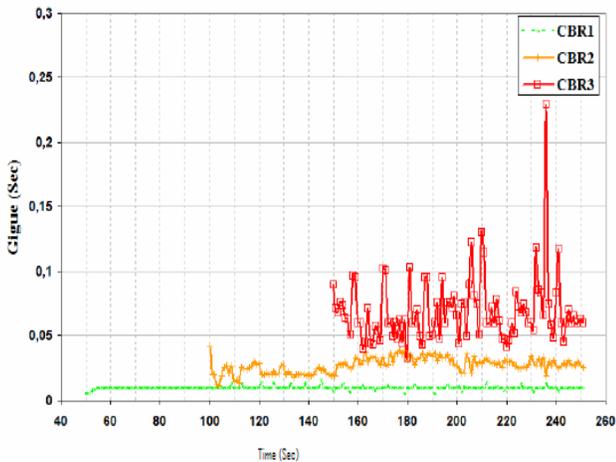

(a) Jitter using three different flows CBR / UDP: case of DIFS

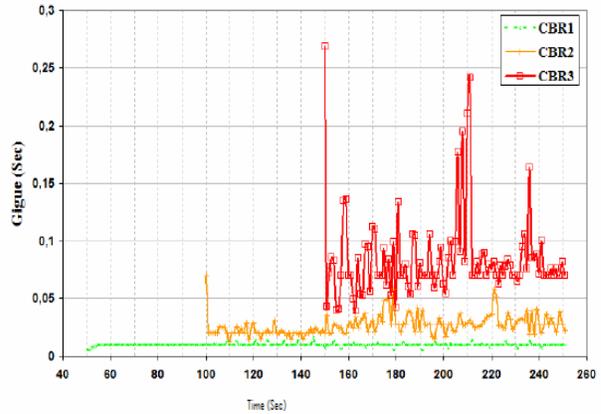

(b) Jitter using three different flows CBR /UDP: case of DF-DCF

Figure 4. Jitter introduced by the MAC layer: (a) DIFS, (b) DF-DCF

This last point can have fewer collisions then fewer retransmissions. Indeed, the mechanism of transmission of IEEE 802.11 based on an exponential binary backoff window may lead to delays and jitter significant and this feature is incompatible with the constraints of real-time applications [18]. We note also that the functioning of DF-DCF allows for the stabilization of the jitter of the flows of highest priority (Figure 4.a/b/). The evaluation that we conducted with UDP flows only, shows a great improvement in the differentiation of services compared to DIFS. This improvement is expressed in terms of reducing delays by flows of highest priority. This is achieved by the loss of frames whose deadline has been reached. Contrary to what one might think, these losses do not impair the overall loss rate of each stream. In fact, our mechanism anticipated only loss of hundreds of frames. This anticipation allows other frames to be quickly served. Indeed, these frames do not keep only less delay in the queues, but also fewer frames will compete for access to the medium.

### B. TCP Traffic

Other interesting results were obtained by replacing CBR/UDP flows with FTP/TCP. In this scenario, each station sends always FTP packets of 1100 bytes. Therefore, they use the TCP transport protocol. The parameters representing the priorities of the three flows for each of the simulated access methods are the same as those used previously for the three UDP flows (see TABLE I). The flow FTP1, the one with the highest priority (with the smallest value $Temax_j$), obtains a very poor quality of service: the value $Temax_j$ is so small that a large number of frames can be lost due to the expiration of their lifetime. Therefore, this flow goes regularly in the phase of Slow-Start and lost frames must be retransmitted by the TCP transport layer. Furthermore, this situation is accentuated when the second terminal begins to transmit: the priority between the two flows is not always visible. When the third terminal begins its transmission, the situation deteriorates to the point that all frames sent by the station STA1, supposed to be the highest priority, and are lost forcing TCP to abort transmission FTP1 after a few seconds. This is due to the values chosen $Temax_j$ which are too low to allow normal operation of TCP. We doubled this value for each flow (TABLE II), but the result remained exactly the same. In fact, larger values of $Temax_j$ give more chance for frames to be transmitted and therefore penalize one TCP flow. Similarly, if we assign a low value to $Temax_j$ for class of service that you want to choose, this will penalize it more rather than give it priority. In other words, more $Temax_j$ is small, the number of frames lost causing increases in TCP retransmissions and decreases radical flow leading to poor differentiation of services. This does not mean that DF-DCF is inadequate to TCP flow. This result demonstrates that only the concepts of priority cannot be the same for UDP flows and TCP flows.

TABLE II. PARAMETERS OF DF-DCF

|  | DF-DCF | |
|---|---|---|
|  | $DIFS_j^{min}$ / $DIFS_j^{max}$ | $Temax_j$ |
| FTP1 | 50µs / 130µs | 300ms |
| FTP2 | 130µs / 210µs | 500ms |
| FTP3 | 210µs / 290µs | 700ms |

UDP is usually used by real-time with the time constraints, but often not very sensitive to losses. Therefore, the delay may force the MAC layer to eliminate frames leading to the suitable choice of the value $Temax_j$. Remember that these losses ultimately lead frame to a decrease in the delay experienced by the flow priority. Then, the idea for the UDP flows will be: more $Temax_j$ is smaller, the flow is more prior. $DIFS_j^{min}$ and $DIFS_j^{max}$ values must be chosen according to the choice of $Temax_j$. Indeed, if $Temax_j$ is chosen small and $DIFS_j^{min}$ and $DIFS_j^{max}$ are chosen high, this can lead to excessive losses in the class: the choice of $DIFS_j^{min}$ and $DIFS_j^{max}$ high while $Temax_j$ is chosen small increases the period of access to medium for each frame and therefore increases the time waiting in the queue, leading to excessive losses.

On the other hand, TCP flows are generally less sensitive to time, but rather to loss, therefore we must eliminate the frames of a TCP flow only when it is the lowest priority and we want to penalize it. The penalization of frames of TCP flow is to increase their waiting time. This is done through an appropriate choice of parameters $DIFS_j^{min}$ and $DIFS_j^{max}$. Regarding the value $Temax_j$ we can use the rule to take the same $Temax_j$ value for all classes of services. This value must be chosen appropriately so that in case of contention with a flow of more priority, frames lost will be that belonging to the lower priority flows leading to a differentiation of services strictly. Depending on the rules that we established earlier, we have assigned new parameters to different classes of services (see TABLE III). As for UDP flows, the simulation results of DF-DCF are compared to those of DIFS.

TABLE III. PARAMETERS OF DF-DCF AND DIFS

|  | DF-DCF | | DIFS |
|---|---|---|---|
|  | $DIFS_j^{min}$ / $DIFS_j^{max}$ | $Temax_j$ | $DIFS_j$ |
| FTP1 | 50µs / 130µs | 375ms | 50µs |
| FTP2 | 130µs / 210µs | 375ms | 130µs |
| CBR3 | 210µs / 290µs | 375ms | 210µs |

Figure 5 shows a better differentiation of services between the TCP flows. In addition, the priorities between flows are more stringent than those obtained with DIFS. Indeed, we see very well in Figure 5.b. that unlike the DIFS differentiation (Figure 5.a), the RTT is clearly differentiated between the three flows. The explanation for this clear improvement, by DF-DCF, lies in the evolution of contention window for each TCP flow. Indeed, we note that the lowest priority flow remains mostly in the Slow-Start phase. This can restrict the flow compared to other flows. FTP3, with a lower bit rate, presents fewer constraints. This causes an improvement in the RTT for the other two flows. Improvement of RTT leads in turn to improve the flow. As for the DIFS differentiation, differentiation of services between flows is almost invisible (Figure 5.a) because, due to its adaptive behavior, no loss is observed by TCP. Indeed, TCP is designed to fit the available flow. Therefore, through the mechanisms of MAC level retransmissions of any lost frame due to collision is transmitted directly by the MAC layer. This was broadcast before the TCP timer expires; no loss is felt at the level of TCP. With each new arrival of a flow, the rate available for each flow in the radio link reduces, causing a decrease in the slope of the congestion window of TCP.

However, this decrease in slope does not prevent the congestion window to continue to grow less rapidly, but it continues to grow since no loss is indicated to it by the absence of a TCP-ACK. The decrease in the slope of the congestion window is due, in turn, to increased RTT which is caused by the arrival of a new flow. Mitigation of service differentiation between TCP flows when DIFS differentiation is used, is due to that this mechanism avoids the losses at the transmitter. In our mechanism, these losses are introduced through the timers $Temax_j$. This no longer means the same as for real-time. We note finally, as in the case of UDP flows, that no loss of total flow is felt here, because all that is lost by the lower priority flows is won by the flow of highest priority.

(a) RTT using three different FTP/TCP flows: case of DIFS

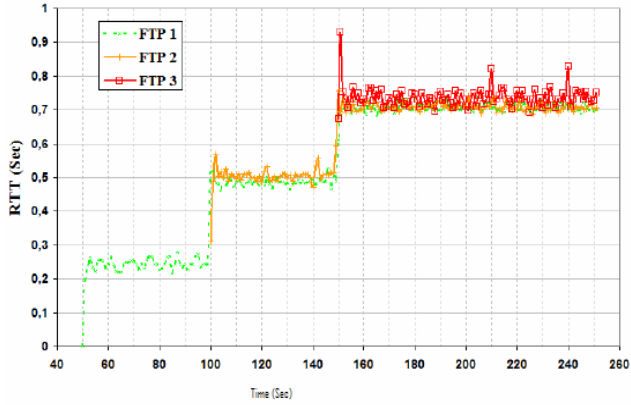

(b) RTT using three different FTP / TCP flows: case of DF-DCF

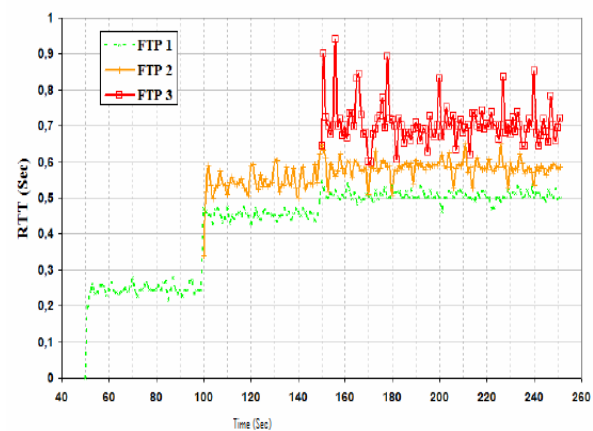

Figure 5. RTT introduced by the MAC layer: (a) DIFS, (b) DF-DCF.

The evaluation of performance that we have achieved with TCP traffic only, allows us to deduce in the first instance a rule for the choice of parameters for differentiation of services. Indeed, a bad choice of the parameter $Temax_j$ may strongly lead a TCP flow into excessive eliminations of frames. We have therefore concluded that the parameter $Temax_j$ must be chosen high in order not to disadvantage the flows of highest priority. We also take the same value $Temax_j$ for all classes of services. The choice of parameters $DIFS_j^{min}$ and $DIFS_j^{max}$ can then set the priority of each of them. This rule is therefore different from that measured for UDP flows. Indeed, UDP is usually used by flows in real time constraints. Therefore, the smaller will be the value $Temax_j$, the highest priority class of service will be. The choice of parameters $DIFS_j^{min}$ and $DIFS_j^{max}$ will then be taken depending on the choice made for $Temax_j$. The performance using this rule allows us to demonstrate the effectiveness of the differentiation of services offered by DF-DCF compared to that obtained by DIFS. This is achieved through the loss of some frames belonging to the lower priority flows and TCP retransmissions. These losses are forcing TCP to reduce the throughput of the corresponding flow (the lower priority). Throughput lost by those flows will be earned by the highest priority flows and the improvement of their RTT.

### C. Mixing UDP and TCP traffic

Having demonstrated the effectiveness of DF-DCF for the differentiation of services between the flows between the UDP and TCP flows, we propose in this section to analyze the behavior of DF-DCF towards a mix of UDP and TCP traffic. Since UDP is generally used for real-time, we therefore give priority to UDP flows compared to TCP flows. Therefore, we will focus on analyzing the QoS assigned to UDP flows when they come into competition with TCP flows. To this end, we have simulated the following scenario: a flow FTP/TCP and two flows CBR/UDP.

TABLE IV.  PARAMETERS OF DF-DCF AND DIFS.

|  | DF-DCF | | DIFS |
|---|---|---|---|
|  | $DIFS_j^{min}$ / $DIFS_j^{max}$ | $Temax_j$ | $DIFS_j$ |
| **CBR3** | 50μs / 130μs | 150ms | 50μs |
| **CBR2** | 130μs / 210μs | 250ms | 130μs |
| **FTP1** | 210μs / 290μs | 1s | 210μs |

The three flows generated are denoted FTP1, CBR2 and CBR3. FTP1 is with the lowest priority, the highest priority going to CBR3 and CBR2 with a medium priority. The parameters representing the level of service for each of the access methods used are illustrated in TABLE IV. Note that the value $Temax_3$ corresponding to flow FTP1 is chosen sufficiently large (1s) so that the TCP flow is not too disadvantaged compared to UDP flows.

Figure 6.b shows clearly that the delay of the flow FTP1 increases whenever a new CBR flow is initiated leading to a clearer differentiation of services with DIFS. Indeed, we see clearly in Figure 6.a no differentiation is achieved between the flows FTP1 and CBR2 when the flux CBR3 begins its transmission. DF-DCF can also reduce delays in the three flows (Figure 6.b) compared with those obtained by DIFS (Figure 6.a). We also note that the flows CBR have the same quality of service when they are competing

with/without a TCP flow. Indeed, the time CBR flows are still guided by the deadline $Temax_j$. This is achieved with an increase in both the flow FTP1 period and changes of this period, this increase are not critical to the operation flow FTP1. The results obtained with TCP traffic only were confirmed in this section also. Indeed, evaluation of performance we have achieved here is a clear service differentiation between different flows. Unlike a DIFS differentiation, the existence of TCP traffic does not affect the performance obtained by the UDP traffic in DF-DCF.

(a) Delay using flow FTP/TCP and two different flows CBR/UDP: case of DIFS

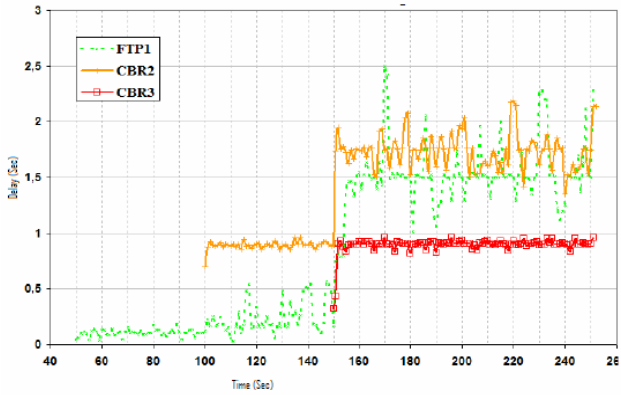

(b) Delay using flow FTP/TCP and two different flows CBR/UDP: case of DF-DCF

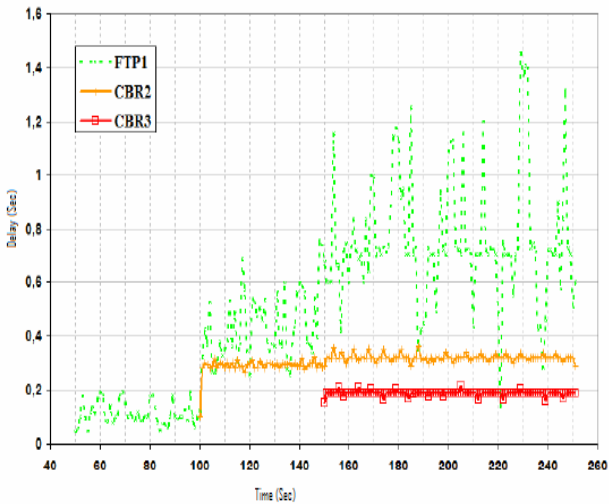

Figure 6. Delays introduced by the MAC layer: (a) DIFS, (b) DF-DCF

## IV. CONCLUSION AND FURTHER WORK

In this article, we discussed the main approaches proposed for the support of the quality of service in IEEE 802.11 networks. This study has identified a number of findings: on the one hand, complexity of management introduced by the centralized access methods that make them hardly viable in practice and on the other hand, inefficiency in some requirements for distributed mechanisms of differentiation of services. A new DF-DCF mechanism was suggested for a differentiation of services effective for all flows. It is based on an extension of the DCF access method applied to the IEEE 802.11 network. The DF-DCF method associated to each frame a level of service depending on their time waiting in the queue like the EDF scheduling policy. This mechanism also has the advantage of being distributed by definition, thus avoiding overload due to the signaling induced by centralized methods. The performance of DF-DCF gives very promising results. Indeed, we have validated by simulation that the differentiation of services between flows belonging to different classes of service, is clearly improved compared to a simple differentiation DIFS. Unlike the other mechanisms, differentiation of services offered by DF-DCF is effective regardless of the type of traffic traversing the network: UDP flow, TCP connection or a mix of UDP and TCP flows. This differentiation of services is particularly achieved through a judicious choice of parameters representing the classes of services. On the other hand, since TCP is generally used by flows more tolerant times but very sensitive to losses, our approach tries to reduce the rate of elimination of frames belonging to a TCP flow. One last very important feature of DF-DCF is that of rates achieved. The DF-DCF method seems to have the same performance in terms of saturation throughput, than other methods of distributed access [8, 9, 17]. At the same time, a control algorithm enables the sharing of bandwidth available in each class should be used to maintain the loss rate below a certain threshold. A permissible loss rate helps maintain a satisfactory level of real-time applications quality. Among the further works, the extension of this work by the study of the scalability seems very interesting to identify the influence of number of nodes and mobility of some of the proposed solution. In addition, the performance evaluation of the QoS concept is under study after its effective implementation in a real mobile environment.

In addition, we suggest coupling mechanisms for service differentiation at the MAC and network layers to build and secure a scheme of QoS from end to end. A research topic that remains open and is a direct result of our work is to optimize the intercellular handover to minimize its negative effects on the differentiation of services. A good approach to managing the handover will reduce the time additional leading to broken communication and rate of loss. This mechanism of QoS will be more suitable for Voice over IP communication.

## REFERENCES

[1] Broadband Radio Access Networks (BRAN), "High Performance Radio LAN (Hiperlan) Type 1: functional

Specification", Project ETSI BRAN of standardisation, EN 300 652, July 1998.
[2] Broadband Radio Access Networks (BRAN), "High Performance Radio LAN (Hiperlan) Type 2: System Overview", Project ETSI BRAN of standardisation, TR 101 683, April 2000.
[3] LAN MAN Standards of the IEEE Computer Society, "Wireless LAN medium access control (MAC) and physical layer (PHY) specification", IEEE Standard 802.11, 1997.
[4] Web Site of Wifi Alliance, "www.wi-fi-.org".
[5] S. Kapp, "802.11 : leaving the wire behind", IEEE Internet computing, Vol. 6, No. 1, pp. 82-85, January/February 2002.
[6] S. Kapp, "802.11a. More bandwidth without wires", IEEE Internet computing, Vol. 6, No. 4, pp. 75-79, July/August 2002.
[7] IEEE 802.11 Task Group e, "Draft Supplement to IEEE Std 802.11 – Part 11 : Wireless MAC and physical Layer Specifications : MAC enhancements for quality of service", 2002.
[8] I. Aad and C. Castelluccia, "Differentiation mechanisms for IEEE 802.11", IEEE Infocom'01, Anchorage, Alaska, April 2001.
[9] I. Aad and C. Castelluccia, "Priorities in WLANs", Computer Networks, Vol. 41, No. 4, pp. 505-526, March 2003.
[10] A. Veres, A.T. Campbell, M. Barry and L.H. Sun, "Supporting Service Differentiation in Wireless Packet Networks using Distributed Control", IEEE Journal of Selected Areas in Communications (JSAC), Vol. 19, No. 10, pp. 2094-2104, October 2001.
[11] NH. Vaidya, P. Bahl and S. Gupa, "Distributed fair scheduling in a wireless LAN", In the sixth Annual International Conference on Mobile Computing and Networking, Boston, USA, August 2000.
[12] A. Lindgren, A. Almquist and O. Shelen, "Quality of Service Schemes for IEEE 802.11 Wireless LANs – An evaluation", ACM/Kluwer Journal Of Special Topics in Mobile Networking and Applications (MONET), Vol. 8, No. 3, pp. 223-235, June 2003.
[13] I. Niang, B. Zouari, H. Afifi and D. Seret, "Amélioration de schémas de QoS dans les réseaux sans fil 802.11", Colloque Francophone sur l'Ingénierie des protocoles, Montréal, Canada, CFIP'02, May 2002.
[14] Y.M. Gh. Doudane, R. Naja, G. Pujolle, and S. Tohme, "P3-DCF : Service Differentiation in IEEE 802.11 WLANs using Per-Packet Priorities", IEEE semi-annual Vehicular Technology Conference, VTC'03-Fall, Orlando, Florida, October 2003.
[15] L. Kleinrock, "Queueing Systems, Volume II: Computer Applications", John Wiley and sons. Wiley Interscience, New York, 1976.
[16] Network Simulator, NS-2.lb9,
http ://www.isi.edu/nsnam/ns/
[17] G. Bianchi, "Performances Analysis of the IEEE 802.11 Distributed Coordination Function", IEEE Journal of Selected Areas in Communications (JSAC), Vol. 18, No. 3, pp. 535-547, March 2000.
[18] W.-T. Chen, B.-B. Jian and S.-C. Lo, « An adaptive retransmission scheme with QoS support for the IEEE 802.11 MAC enhancement », IEEE Vehicular Technology Conference Spring 2002, IEEE VTC Spring'02, (Birmingham, Alabama), May 2002.